\documentclass[a4paper,11pt]{article}
\pdfoutput=1
\usepackage{jcappub}
\usepackage{graphics,graphicx, color}
\usepackage{dcolumn, amsmath,amssymb}

\newcommand{\be}{\begin{equation}}
\newcommand{\ee}{\end{equation}}
\newcommand{\ba}{\begin{eqnarray}}
\newcommand{\ea}{\end{eqnarray}}
\newcommand{\bi}{\begin{itemize}}
\newcommand{\ei}{\end{itemize}}

\newcommand{\yue}{}

\begin{document}

\title{A Strong Test of the Dark Matter Origin of a TeV Electron Excess \\Using IceCube Neutrinos}

\author[a,b]{Yue Zhao\footnote{zhaoyhep@umich.edu}}
\affiliation[a]{Tsung-Dao Lee Institute, and Department of Physics and Astronomy, Shanghai Jiao Tong University, Shanghai 200240}
\affiliation[b]{Leinweber Center for Theoretical Physics, University of
Michigan, Ann Arbor, MI 48109}
\author[c, d]{Ke Fang\footnote{kefang@umd.edu}}
\affiliation[c]{University of Maryland, Department of Astronomy, College Park, MD, 20742}
\affiliation[d]{Joint Space-Science Institute, College Park, MD, 20742}
\author[e]{Meng Su\footnote{mengsu.astro@gmail.com}}
\affiliation[e]{Department of Physics and Laboratory for Space Research,
the University of Hong Kong, PokFuLam, Hong Kong SAR, China}
\author[c, d]{M. Coleman Miller\footnote{miller@astro.umd.edu}}

\abstract{ {\yue{Due to the electroweak symmetry, high energy
neutrinos and charged leptons are generically produced
simultaneously in heavy dark matter decay or annihilation process.
Correlating these two channels in dark matter indirect detections
may provide important information on the intrinsic production
mechanism. In this paper, we demonstrate this point by studying the
tentative excess in the electron spectrum at 1.4 TeV reported by the
DArk Matter Particle Explorer (DAMPE).}} A non-astrophysical
scenario in which dark matter particles annihilate or decay in a
local clump has been invoked to explain the excess. If $e^\pm$
annihilation channels in the final states are mediated by
left-handed leptons as a component in the $SU(2)_L$ doublet,
neutrinos with similar energies should have been simultaneously
produced.  We demonstrate that generic dark matter models can be
decisively tested by the existing IceCube data. In case of a
non-detection, such models would be excluded at the $5\sigma$ level
by the five-year data for a point-like source and by the ten-year
data for an extended source of dark matter particles with
left-handed leptons. This serves as an example of the importance of
correlating charged lepton and neutrino channels.  It would be
fruitful to conduct similar studies related to other approaches to
the indirect detection of dark matter. }

\maketitle

\section{Introduction}
{\yue Correlating various channels in dark matter indirect
detections is crucial to reduce the astrophysical uncertainties and
extract information from the dark matter annihilation or decay
processes.}

{ \yue TeV cosmic ray electrons (CRE) and neutrinos provide ideal
probes of potential signatures of dark matter particle annihilation
or decay in the vicinity of the solar system. TeV electrons cool
very fast while propagating in the Milky Way, thus the sources of
such high energy electrons must be within 1 kpc.  The direction of
the source is largely erased in the electron flux. On the other
hand, the propagation of neutrinos is barely affected and provides
excellent directional information. Furthermore, high energy
neutrinos and charged leptons are generically correlated with each
other due to the electroweak symmetry. If a hint of an excess is
suggested in one channel, searching for the corresponding signal in the other channel can solidly test such a hypothesis.}

The CRE spectrum has recently been directly measured up to 5 TeV
using the spaceborne DArk Matter Particle Explorer (DAMPE;
\cite{2017APh....95....6C}).  The DAMPE measurement of the CRE
spectrum has unprecedented high energy resolution, low background,
and well controlled instrumental systematics. Although the majority
of the spectrum can be fitted by a smoothly broken power-law model
with a spectral break at E$\sim$0.9 TeV, a tentative peak at
$\sim$1.4 TeV in $e^+e^-$ total spectrum has been claimed
\cite{Ambrosi:2017wek}. The excess of $e^+e^-$ pairs at 1.4 TeV is
approximately $2.5\times10^{-8}~\rm{GeV^{-1}~s^{-1}~sr^{-1}~m^{-2}}$
compared with the best continuum fit of the electron-positron energy
spectrum of DAMPE \cite{Yuan:2017ysv}. Structures around TeV are
also evident in the electron-positron spectrum measured by the
Calorimetric Electron Telescope (CALET), although more statistics
and refined data analysis are needed to reach a conclusion
\citep{2017PhRvL.119r1101A}.

A sharp peak in the $e^+e^-$ spectrum is hard to explain with a distant source.  Electrons and positrons with TeV energy quickly lose energy via synchrotron radiation and inverse Compton scattering processes. Even if the initial spectrum is monoenergetic, propagation introduces a dispersion in the energy. If the peak is confirmed by future data, the  $1.5\%$ energy resolution of DAMPE  in the TeV regime would require that  the source of the TeV electrons and positrons  has to be within 0.3 kpc \cite{Yuan:2017ysv,2017arXiv171200370G,2017arXiv171200037C,2017arXiv171111579L,2017arXiv171111563D,  2017arXiv171111452C,2017arXiv171111376A,2017arXiv171111333G,2017arXiv171111182C,2017arXiv171110995F,2017arXiv171200362J,   2017arXiv171200011C, 2017arXiv171200005H, 2017arXiv171111052Z, 2017arXiv171110989Y,  2017arXiv171110996F, Yang:2017cjm}.

Two types of scenarios have been proposed to explain the DAMPE electron-positron flux. In an astrophysical scenario, an isolated young pulsar could produce such a sharp peak if it rotates relatively slowly and has a mild magnetic field (e.g., \citep{Yuan:2017ysv, 2017arXiv171200011C}). In a non-astrophysical scenario, small dark matter (DM) substructure, such as a nearby clump, can produce a large $e^+e^-$ flux due to its enhanced DM density (e.g., \citep{2017arXiv171111563D, 2017arXiv171111579L}). It is thus crucial to explore correlated multi-messenger signals that may help to discriminate between possible explanations \citep{2017arXiv171200370G}. If an association is established, the directional information carried by weakly-interacting particles such as neutrinos will be crucial to finding the source.

From a particle physics viewpoint, dark matter models fall into two
generic groups according to the final state electron chirality. The
left-handed electrons are fundamentally different from right-handed
electrons. If the $e^+e^-$ produced by DM are left-handed, one
generically expects a comparable neutrino flux simultaneously. This
is because the electroweak symmetry breaking (EWSB) scale is at
O(100) GeV, thus induces negligible difference between left-handed
electrons and neutrinos at an energy as high as 1.4 TeV. The
associated neutrinos should be almost monochromatic and may carry
important directional information if they come from a nearby DM
clump.

In this work, using the $e^\pm$ flux at 1.4 TeV measured by DAMPE as
a reference of the TeV electron excess, we illustrate that searching
for associated neutrino signals can provide strong tests on DM
interpretations. We first demonstrate that the accompanying
neutrinos, with a monochromatic energy of 1.4~TeV, are naturally
expected in a generic class of models. Then we briefly discuss the
flavors of neutrinos when they reach the earth. Based on the
neutrino background levels in the IceCube one-year public data, we
demonstrate that the existing $\sim$8-year IceCube data is
sufficient to decisively test the possibility that the DAMPE $e^\pm$
excess is left-handed and produced from the annihilation or decay of
DM.

{\yue We emphasize that, in this paper, we use the tentative peak
appearing in DAMPE's measurement as a demonstration to show the
benefit on studying the correlation between electron and neutrino
channels. We are not limited by this particular choice of benchmark
scenario and similar studies can be applied in more general cases.}

\section{Classification of Models}
If a left-handed electron is involved
in the DM annihilation process, generically a comparable neutrino
flux is also generated due to $SU(2)_L$ symmetry. We present a
classification of models and demonstrate that a neutrino flux is
naturally produced. To quantify the relative ratio, we define
$\eta_i$ as
 \begin{align}\label{eqn:eta}
\eta_i=L_{\nu_i}/L_{e} \end{align} where $L$ is the injection
luminosity and $i$ labels the lepton flavor. We do not distinguish
$e^+$ and $e^-$ because they are indistinguishable using DAMPE at
high energy. Similarly, IceCube data cannot distinguish neutrinos
from anti-neutrinos. We have not yet included the effects of
neutrino oscillation. If only electrons and positrons are produced,
$\eta_{\mu,\tau}=0$. For the flavor universal scenario,
$\eta_{e,\mu,\tau}$ are all equal.

There are two ways to link dark matter with a standard model (SM)
lepton: s or t-channel exchange of a mediator. $Z'$ is a typical
s-channel mediator, which is the gauge boson of a new $U(1)$ gauge
group. Label the charges of left and right-handed leptons as
$q'_{L,i}$ and $q'_{R,i}$, $\eta_i$ is
$\eta_i={{q'}^2_{L,i}}/{\left({q'}^2_{L,e}+{q'}^2_{R,e}\right)}$. If
$Z'$ does not couple to right handed leptons, then $\eta_i$ is 1.
With comparable charge assignments, $\eta_i$ is O(1). In
\cite{Athron:2017drj}, the s-channel models were studied and we
discuss the classification of the t-channel models in this section.

{\it (i) $SU(2)_L$ singlet.} If dark matter is a singlet of
$SU(2)_L$ and the leptons in the final states are a $SU(2)_L$
doublet, the mediator should be an $SU(2)_L$ doublet,
\begin{align}
L \supset \lambda_{i} \phi \Psi_{L,1} L_i + M_{\Psi}
\Psi_{L,1}\Psi_{L,2} + h.c.
\end{align}
$\Psi_{L,1}$ and $\Psi_{L,2}$ are heavy vector-like fermions in the
fundamental representation of $SU(2)_L$ with hypercharge +1 and -1.
DM annihilation is mediated by $\Psi$. EWSB may introduce a mass
splitting between $SU(2)_L$ components of $\Psi$. However EWSB
happens around 100 GeV, much smaller than the hypothesized 1.4~TeV
DM mass. The induced mass splitting is small, $<10\%$, and
comparable neutrino flux should be produced, i.e. $|1-\eta_e|< 0.1$.

If DM is a real scalar, the leading annihilation is p-wave due to
chirality suppression. Comparing with complex scalar DM, to achieve
the same $e^+e^-$ production rate, the DM energy density in the
clump needs to be much larger, $O(\sim 10^3)$, assuming similar
coupling constants and $M_{\Psi}$. DM particle may be fermionic. For
a gauge singlet, it can pair with itself and form a Majorana
fermion.  DM can also be a Dirac fermion composed of two Weyl
spinors. For Majorana DM, the dominant annihilation channel is again
p-wave suppressed.

{\it (ii) Non-trivial representation under $SU(2)_L$.} Let us first
consider the fundamental representation; we comment on higher
representations later. If a DM particle is an $SU(2)_L$ doublet, the
heavy mediator transforms under trivial or adjoint representation.
We introduce two sets of Weyl spinors, $\chi_1$ and $\chi_2$, which
are $SU(2)_L$ doublets with hypercharge +1 and -1. With a
vector-like mass, their EM neutral components form a Dirac fermion
as DM.

If the mediators,  $\phi_n$ and $\phi_c$, are $SU(2)_L$ singlets.
They carry 0 and +2 hypercharge respectively. The Lagrangian is
\begin{equation}
L \supset \lambda_{1,i}\chi_1 L_i \phi_n + \lambda_{2,i}\chi_2
L_i\phi_c\nonumber+M_\chi \chi_1\chi_2
+m_n^2\phi_n^2+m_c^2\phi_c^2 +h.c.
\end{equation}
$M_\chi$ determines the DM mass scale.  The relative ratio is
determined by $\lambda_{1,i}$, $\lambda_{2,i}$ and $|m_{n,c}|$. With
generic choices, $\eta_e\sim O(1)$.

$\eta_e$ is a free parameter here because no symmetry relates
$\lambda_{1,i}$, $\lambda_{2,i}$ and $|m_{n,c}|$. Such freedom is
gone if the heavy mediator transforms non-trivially under $SU(2)_L$,
e.g., an adjoint representation when DM is a doublet or when DM is
in higher representation of $SU(2)_L$. Then $\eta_e$ is expected to
differ from unity by at most 10\%.

In the $SU(2)_L$ doublet case, DM cannot be a Dirac fermion, because
of strong dark matter direct detection constraints. Similar to the
higgsinos in Minimal Supersymmetric Standard Model, a small mixing
after EWSB with other fermions, such as the Wino or Bino, breaks a
Dirac fermion into two Majorana fermions. The DM direct detection
constraints can then be evaded. Since DM is effectively Majorana,
its annihilation is p-wave suppressed.

We now comment briefly on the DM decay which can be realized when
mediators are lighter than DM. {\yue The lifetime of the dark matter
particle can be cosmologically long if the coupling between the DM
field and the mediator is very weak. This can be naturally realized
if such vertex violates an approximate global symmetry.} The energy
of the electron and neutrino is
 \begin{align}
E_{e,\nu} = \frac{m_{DM}^2-m_{med}^2}{2m_{DM}}
 \end{align}

Some components of the mediator are charged, but the collider
constraints are weak. If the heavy charged particles are long-lived,
their mass can still be O(100) GeV
\cite{ATLAS:2014fka,Khachatryan:2016sfv}. The charged heavy
particles may lose its charge by decaying to charged SM leptons and
neutral particles. This evades potential problems in cosmology. The
charged leptons from the secondary decay is softer and can hide in
the continuous cosmic-ray background.

{\bf Neutrino Oscillation} The flavor of a neutrino changes during
propagation. We present the detailed calculation for neutrino
oscillations in the Appendix. For a sizable DM clump considered
here, i.e., O(10) pc,  the observed neutrino fraction, $\kappa_i$,
is independent on the DM clump size and the distance to us  as shown in Figure~\ref{fig:eleconly}. In the
Appendix, we show the flavor fraction in two scenarios,
electron-only and flavor universal.  The undetermined CP violating
angle in PMNS matrix $\delta_{CP}$ only affects the results by O(1),
thus it does not change our conclusion qualitatively.

\section{Neutrino Flux}
After leaving a source, TeV electrons diffuse in the Galactic magnetic field while neutrinos travel in a straight path to reach the earth. We now describe how the electron and the neutrino fluxes are connected.

\begin{figure}[t]
\begin{center}
\includegraphics[width=.7\textwidth]{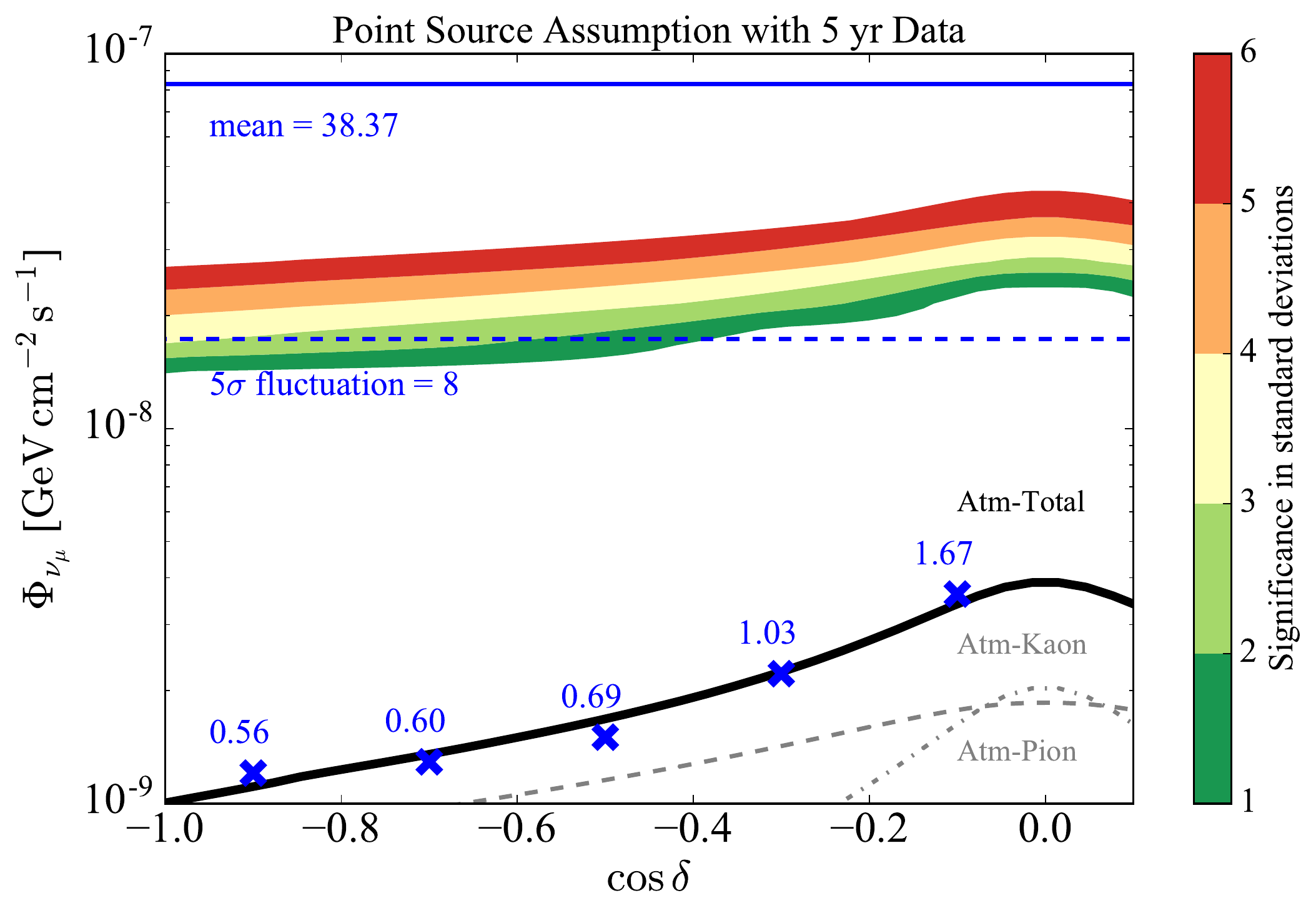}
\includegraphics[width=.7\textwidth]{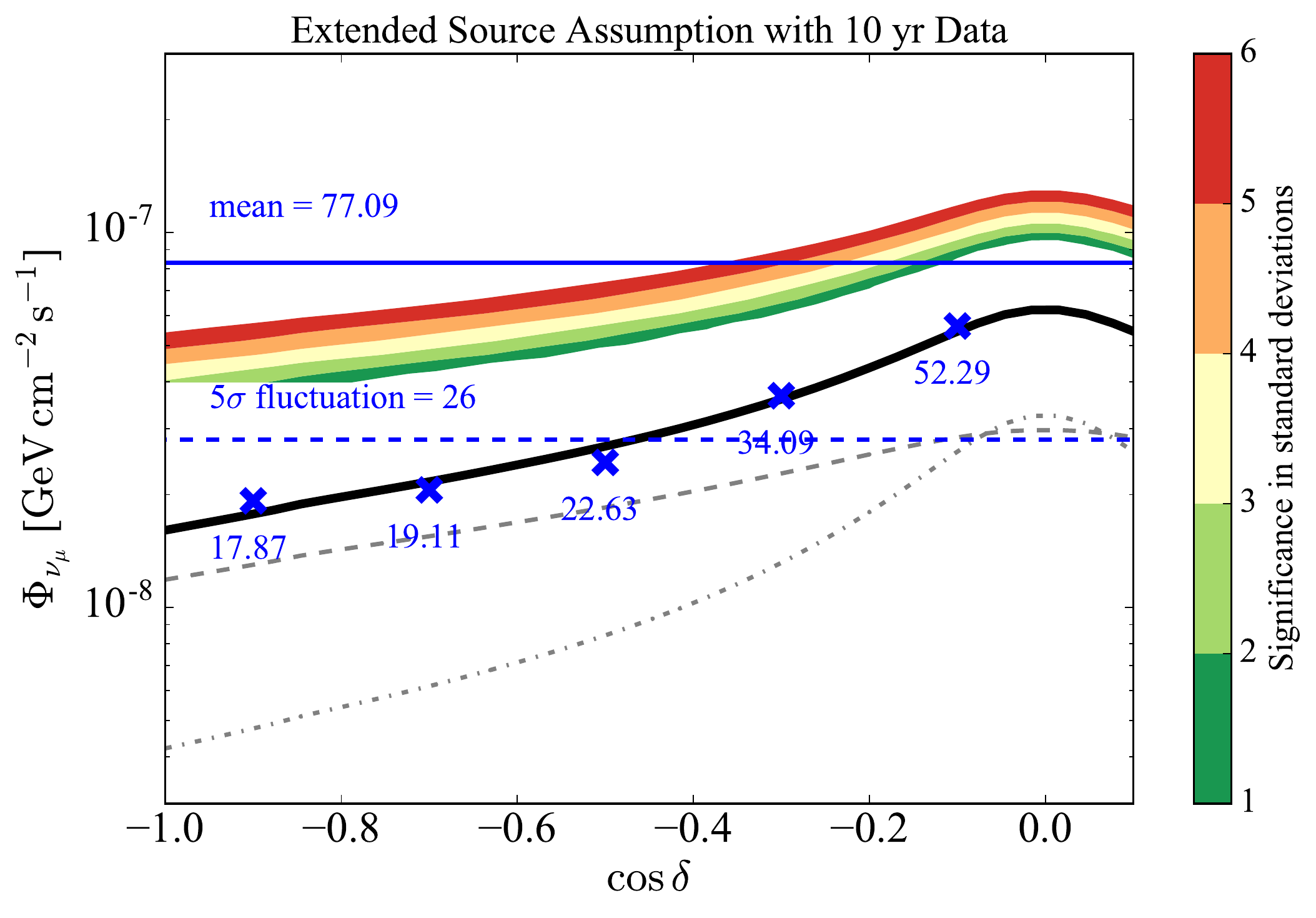}
\end{center}
\caption{The flux of neutrinos from a fiducial dark matter source
that may produce {\yue the 1.4 TeV excess} in the electron spectrum
measured using DAMPE (mean and a hypothetical 5$\sigma$ downward
fluctuation in blue lines), compared with the flux of the
atmospheric neutrino background around 1.4 TeV (kaon and pion
components are indicated by grey dash-dotted and dashed lines, and
the total is shown by the black solid line). The top and bottom
panels assume that the source is point-like and extended
respectively. In the top (bottom) panel, the blue cross data points
present the average number of TeV events in a
$0.5^\circ\times0.5^\circ$ ($2^\circ\times2^\circ$) sky patch in the
five-year (ten-year) IceCube data, which is obtained by scaling that
in the IceCube-86 2011-2012 data by the corresponding number of
years \cite{2016PhRvL.117g1801A}. The colored contours show the flux
needed for the source to stand out of the atmospheric background
with 1 to 6$\sigma$ significance levels. We find that the existing
$\sim$8-year IceCube data can decisively test a generic class of
dark matter models that produce neutrinos and the DAMPE TeV
electrons simultaneously.   } \label{fig:excess}
\end{figure}

The density of particles $n$ at a location $\vec{x}$ from a source at $\vec{x}_s$ follows the transport equation \citep{1990acr..book.....B}
\begin{equation}
    \frac{\partial n(\vec{x}, E)}{\partial t} = D(E)\,\nabla^2\,n+Q_s(E)\delta(\vec{x} - \vec{x}_s)\,,
\end{equation}
assuming that there is no energy loss and that the source is stationary.
Here $D(E)=D_0\,\left(E/E_0\right)^\alpha$ is the spatial diffusion coefficient, with $D_0=10^{28}\,\rm cm^2\,s^{-1}$, $\alpha=1/3$ and $E_0=3$~GeV corresponding to the diffusion coefficient in the interstellar medium \citep{1990acr..book.....B}, and $Q_s(E)$ is the particle injection rate.
The solution can be written as \footnote{When there is more than one DM clump nearby, our calculation would still apply if a single clump makes the dominant contribution to the electron and neutrino fluxes. }
 \begin{align}
    n_e(E) = \frac{Q_e(E)}{4\pi\,R_s\,D(E)}
 \end{align}

Assuming that the DAMPE excess is due to a  monochromatic electron population  with an observed energy density $w_e = n_e\,E_e= 9.8\times10^{-19}\,\rm erg\,cm^{-3}$ \citep{Yuan:2017ysv}, the electron injection power is $L_e = 7.6\times10^{32}\,\left({R_s}/{0.2\,\rm kpc}\right)\left({D(E)}/{10^{29}\,\rm cm^2\, s^{-1}}\right)\,\rm erg\,s^{-1}$.

A population of neutrinos is produced at the same time as the
electrons. The injection powers of the two species are connected by
Eq.~\ref{eqn:eta}. The neutrino flux is then
\begin{eqnarray}\label{eqn:F_nu}
F_{\nu_j} &=&   \frac{\sum_i\,\eta_i\,\kappa_j\,L_e}{\delta \left(\log E_\nu\right)\,4\pi\,R_s^2} \\
&=& 8.3\times10^{-8}\,\eta\,\left(\frac{R_s}{0.2\,\rm kpc}\right)^{-1}\left(\frac{D(E)}{10^{29}\,\rm cm^2\, s^{-1}}\right) \nonumber \\
&&\left(\frac{\sigma\left(\log
E_\nu\right)}{1.2}\right)\left(\frac{\sum_i\,\eta_i\,\kappa_j}{1}\right)\,\rm
GeV\,cm^{-2}\,s^{-1} \nonumber \\ \nonumber
\end{eqnarray}
where
$\delta\left(\log E_\nu\right)$ is the energy resolution of neutrino
events, which is $\delta\left(\log E_{\nu_\mu}\right)\sim 1.2$ for
the IceCube $\nu_\mu$ events based on the reconstruction of the muon
energy in the IceCube detector   \citep{2016PhRvL.117g1801A}. $\kappa_i$ is the
fraction of neutrino in each flavor at the time of detection. More
details about neutrino oscillation can be found in the Appendix.

The neutrino flux depends on the source distance. A DM clump with a total mass of $10^7-10^8\,M_\odot$ with a distance of $0.1-0.3\,\rm kpc$ has been suggested to account for the DAMPE TeV data \citep{Yuan:2017ysv}. The source is unlikely to be more distant than O(1) kpc, as electrons would have suffered from significant energy loss and not present a feature as narrow as in the DAMPE spectrum. We thus choose $0.2\,\rm kpc$ as a default distance for the following calculations.

For comparison, at 1.4~TeV, the averaged conventional muon neutrino flux per solid angle, including both pion and kaon contributions, is $3\times 10^{-5}\,\rm GeV \, cm^{-2}\,s^{-1}\,sr^{-1}$ \citep{2002ARNPS..52..153G}.  The averaged conventional electron neutrino flux per solid angle is $2\times 10^{-6}\,\rm GeV\,cm^{-2}\,s^{-1}\,sr^{-1}$ \citep{2002ARNPS..52..153G}.
If the observed electrons arrive from a preferred direction, which for example may be indicated by an anisotropy in the electron data, one can search for coincident neutrinos in the associated sky location. The average angular resolution of the IceCube detector is $0.5^\circ$ for $\nu_\mu$ events and a few degrees for $\nu_e$ events at $\sim 1$~TeV. The background flux in the sky patch surrounding the source direction would thus be
$F^{\rm bg}_{\nu_\mu} = 2.3\times10^{-9}\,\left({\delta\theta}/{0.5^\circ}\right)^2\,\rm GeV\,cm^{-2}\,s^{-1}$
and
$F^{\rm bg}_{\nu_e} = 1.5\times10^{-8}\,\left({\delta\theta}/{5^\circ}\right)^2\,\rm GeV\,cm^{-2}\,s^{-1}$

Using the IceCube effective area of $\sim 5000\,\rm cm^2$ at 1.4~TeV, Eq. ~\ref{eqn:F_nu} corresponds to an average of $7.7$ events per year of IceCube data. Therefore it will be easy to use IceCube to identify a bright source if we know its location in the sky.
However, no specific directional information has been provided by the current data of DAMPE \citep{2017APh....95....6C}. This could be due to a limit of the statistics, or because electrons started from a relatively distant location and have lost most of their angular information during a diffusive propagation. Below we investigate the feasibility of a blind search for the associated neutrino signal using the IceCube muon neutrinos events. We do not use cascade events due to their poor angular resolutions, but note that they may be useful for searches for very extended sources as shown in Ref.~\cite{2017arXiv170502383I}.

We analyze the public IceCube data from the full 86-string detector configuration taken during 2011-2012, using only up-going neutrinos from the northern sky to eliminate background muon events \citep{2016PhRvL.117g1801A}. The sample contains a total of 20,145 neutrino events with reconstructed energy in the approximate 320~GeV to 20~TeV range. We select 18,722 events that may have a deposition energy at  1.4~TeV according to a $\delta \left(\log E_{\nu_\mu}\right) \sim 1.2$ energy resolution \citep{2016PhRvL.117g1801A} (between 320~GeV and 3080~GeV). The public data only have the zenith angle of the reconstructed events, rather than the full two-dimensional angular location, so we cannot perform an analysis with these data alone.  In the following, however, we project an analysis based on an extrapolation of the neutrino numbers to a 5-year and a 10-year data set, which we assume will include both the zenith angle and the azimuthal angle for each reconstructed event.

The events are pixelized based on a HEALPix71 \citep{2005ApJ...622..759G} pixelization scheme for spatial binning.
Two source scenarios are considered. In the first scenario, we assume that the dark matter source is point-like.  We choose   a bin size of approximately $0.5^\circ \times 0.5^\circ$ (with the HEALPix parameter $\rm Nside=128$, corresponding to the angular solution of the IceCube $\nu_\mu$ events \citep{2016PhRvL.117g1801A}. In the second scenario, we assume that the source is extended. We choose   a spatial bin size of $\sim 2^\circ\times2^\circ$ ($\rm Nside = 32$), corresponding to  a dark matter clump with a size of $\sim 10\,\rm pc$ and a distance of $\sim 0.2-0.3$~kpc as suggested  for e.g. by the benchmark case in Ref.~\citep{Yuan:2017ysv}. We divide the
data into 6 bins according to the zenith angle of the reconstructed
events, and let all bins have equal solid angles.  We verify that the event number in pixels in all zenith angle bands  follows the Poisson distribution.

Figure~\ref{fig:excess} presents the flux of TeV neutrinos from the
atmospheric background in an element sky patch (a pixel in our analysis) in the point-source and the extended-source scenarios,
comparing to the flux of neutrinos from a fiducial  source that produces the DAMPE TeV electrons.
The
solid black line corresponds to the atmospheric model of
\cite{2002ARNPS..52..153G}, including both  pion and kaon
components. The blue cross data points and their values  show  the average number of events in a pixel at the corresponding  zenith angle expected in  five-year data (top) and ten-year data (bottom). The values are obtained by scaling the average event number in one pixel in the 2011-2012 data by the number of observational years. The dependence of the  event number on the zenith angle is consistent with that of the total   atmospheric neutrino flux. The  scaling between the two is obtained by fitting the data to the model, and its physical meaning is the effective area multiplied by the observational time and the angular size of the pixel.
The colored contours show the flux levels that are needed to reach a 1 to 6 $\sigma$ deviation from a background-only hypothesis. Specifically, the local probability of  deviation is calculated using the Poisson distribution with a mean determined by the average number of atmospheric neutrinos in a pixel. The local probability is then corrected by a trial factor that equals to the number of independent pixels used in the analysis. Finally, the global probability is quoted using the corresponding number of standard deviations for a Gaussian distribution.

The  solid blue line indicates the expected number of neutrino events from a dark matter source that may explain the TeV excess of electrons measured by DAMPE (as described by equation~\ref{eqn:F_nu}). We find that in both point-source and extended source scenarios, the mean flux is high enough to reject a background-only hypothesis with high significance. We additionally  consider a pessimistic scenario in which the number of neutrino events from the dark matter source is 5-$\sigma$ below its mean value, integrated over the uncertainty of the DAMPE electron flux at 1.4~TeV, as indicated by the dashed blue line in both plots. Even in this case, the five-year data is able to reveal a dark matter source in most part of the northern sky with high confidence levels.

\section{Conclusion}
In this letter, we demonstrate the feasibility of testing TeV dark
matter models by neutrino observations. We focus on a benchmark
scenario in which the tentative TeV electron excess measured by
DAMPE is explained by the DM annihilation or decay in a nearby
subhalo. In a generic class of dark matter models where $e^\pm$ from
DM annihilation or decay are in a $SU(2)_L$ doublet, neutrinos of
comparable flux are simultaneously generated.  We have shown that
the existing $\sim$8-year IceCube data is sufficient to identify the
associated neutrinos with high significance, and decisively test any
dark matter models using left-handed leptons to explain the DAMPE
TeV {\yue peak}.

For a point-like subhalo in the northern sky, our results are robust even in a pessimistic scenario where due to fluctuations the neutrino flux is $5\sigma$  lower than expected. For an extended source, ten years of IceCube data can reveal its existence in most parts of the northern sky. But if the source is much more extended than an angular size of  $\sim 2^\circ$, the prospects for using IceCube to detect or constrain the dark matter source would not be as good.  We do note that there are two refinements to our analysis that could improve those prospects significantly.  First, we assumed for simplicity that the neutrino flux from an extended source would be uniform over the whole solid angle.  In reality, the flux would be centrally concentrated in a way specific to the halo properties, so a search for that concentration would yield a stronger signal.  Second, we likewise assumed that the reconstructed background atmospheric neutrino flux is equally spread over the entire $\sim 300-3000$~GeV range, but in fact that flux drops sharply with increasing energy.  As a result, the excess signal at higher reconstructed energies is likely to be substantially larger than in our current conservative estimates.




The propagation of electrons depends on the diffusion coefficient of
the region between the source and the earth. If the  diffusion
coefficient is well below the average value of the ISM ( as it is in
the region surrounding the Geminga pulsar \cite{2017PhRvD..96j3013H,
2017arXiv171106223A}), electrons would be confined to be near the
source for longer. To maintain a non-broadened {\yue peak} feature,
the source would then need to be closer than that in our benchmark
case. Therefore the neutrino flux (in equation~\ref{eqn:F_nu}) does
not strongly depend on the diffusion coefficient.

We have used the IceCube detector to test DM models in this work. Other high-energy neutrino experiments, including the Super-Kamiokande \citep{2016PhRvD..94e2001R} and the ANTARES Telescope \citep{2017arXiv171101496A}  may provide additional sky coverages to find or constrain associated neutrino signals. Future  experiments  such as IceCube-Gen2 \citep{2014arXiv1412.5106I} and  KM3NeT \citep{2016JPhG...43h4001A}   will provide improved sensitivity to examine very extended or extremely faint dark matter halos.

\section*{Neutrino oscillations and injection channels}

\begin{figure}[t]
\begin{center}
\includegraphics[width=0.7\textwidth]{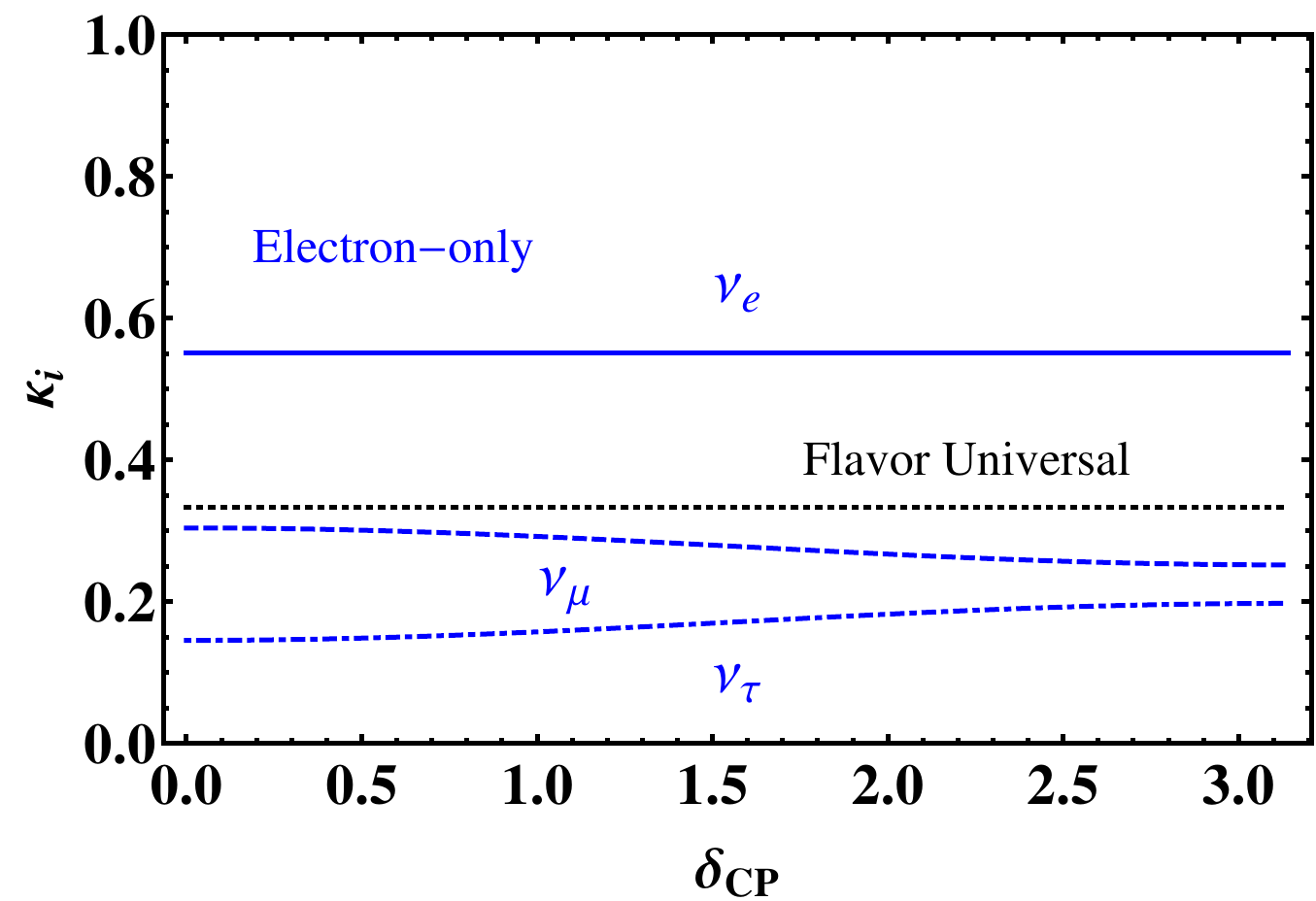}
\caption{ The fraction of each flavor in the measured neutrino flux
around the earth. The blue lines correspond to the scenario in which
only electron neutrinos are produced by DM annihilation. Solid,
dashed and dot-dashed lines are for electron, muon and tau neutrinos
respectively.  The black dotted line is for the flavor universal
scenario, in which the fractions for each flavor is equal to 1/3.}
\label{fig:eleconly}
\end{center}
\end{figure}

Mixing among neutrinos can be characterized by the PMNS matrix, $U$.
Neglecting the possible CP phases from the Majorana mass terms,
there are four parameters in PMNS matrix: three mixing angles and
one CP phase, i.e.
$\{\theta_{12},\theta_{13},\theta_{23},\delta_{CP}\}$. $\theta_{12}$
and $\theta_{13}$ have been measured with good accuracy,
$\textrm{sin}^2(2\theta_{13})=0.093\pm0.008$ and
$\textrm{sin}^2(2\theta_{12})=0.846\pm0.021$. $\theta_{23}$ has
larger uncertainty, $\textrm{sin}^2(2\theta_{23})>0.92$. In the
following calculation, we take the central values of
$\textrm{sin}^2(2\theta_{12})$ and $\textrm{sin}^2(2\theta_{13})$,
while $\textrm{sin}^2(2\theta_{23})$ is taken to be 0.97. The
uncertainties in these values do not change our results
qualitatively.  $\delta_{CP}$ remains to be determined, and we treat
it as a free parameter.

The oscillation of neutrinos is then written as
\begin{eqnarray}
P_{\alpha\to\beta}=&\delta_{\alpha\beta}&-4\sum_{i>j}Re (U^*_{\alpha
i} U_{\beta i}U_{\alpha j} U^*_{\beta j})
\textrm{sin}^2\bigg(\frac{\Delta m_{ij}^2
L}{4E}\bigg)\nonumber\\
&+&2\sum_{i>j}Im (U^*_{\alpha i} U_{\beta i}U_{\alpha j} U^*_{\beta
j}) \textrm{sin}\bigg(\frac{\Delta m_{ij}^2 L}{4E}\bigg)
\end{eqnarray}
Here $\Delta m_{ij}^2$ is the mass square splitting among neutrino
mass eigenstates. In a vacuum, $\Delta m_{21}^2 =
(7.53\pm0.18)\times 10^{-5}\textrm{eV}^2$ and $|\Delta
m_{31}^2|\simeq |\Delta m_{32}^2| = (2.44\pm0.06)\times
10^{-3}\textrm{eV}^2$. The size of the DM clump which generates the monochromatic electron flux can be large, for example O(10) pc as considered in our
benchmark case. For neutrinos with energy O(TeV), this is much larger
than the distance to have one oscillation, i.e.  $\bigg(\frac{\Delta m_{ij}^2 10 \textrm{pc}}{4 \textrm{TeV}}\bigg) \gg 1$.  After averaging the whole clump,

\begin{eqnarray}
\langle
P_{\alpha\to\beta}\rangle=\delta_{\alpha\beta}&-&2\sum_{i>j}Re
(U^*_{\alpha i} U_{\beta i}U_{\alpha j} U^*_{\beta j})
\end{eqnarray}

{(a) Electron-only channel:}
this is the scenario where DM annihilation only produces
electrons and electron neutrinos. The flavor fraction of neutrino flux, $\kappa_i$, are labeled as blue lines in Fig. 1.

{(b) Flavor-universal channel:}
if the DM annihilation is universal in flavor, i.e. the initial flux
for each flavor is the same, the flux observed on earth remains
flavor universal, guaranteed by the unitarity of the mixing matrix. This is represented by the black dotted line in Fig. 1.

\section*{Acknowledgements}
YZ is also supported by US Department of Energy under
grant de-sc0007859. KF acknowledges the support of a Joint
Space-Science Institute prize postdoctoral fellowship at the
University of Maryland. YZ thank the support of grant from the
Office of Science and Technology, Shanghai Municipal Government (No.
16DZ2260200).

\bibliography{DAMPENeu}

\end{document}